\documentclass[12pt]{iopart}
\usepackage{iopams}  
\usepackage{graphicx}
\begin{document}

\title[Fresnel filtering and angular Goos-H\"anchen shift]{Are Fresnel filtering and the angular Goos-H\"anchen shift the same?}

\author{J\"org B. G\"otte$^1$, Susumu Shinohara$^2$ and Martina Hentschel$^{3}$}

\address{$^1$ Max-Planck-Institute for the Physics of Complex Systems, N\"othnitzer Str. 38, 01187 Dresden, Germany}
\address{$^2$ NTT Communication Science Laboratories, NTT Corporation, 2-4 Hikaridai, Seika-cho, Soraku-gun, Kyoto 619-0237, Japan}

\address{$^3$ Technische Universit\"at Ilmenau, Institut f\"ur Physik, Weimarer Str. 25, 98693 Ilmenau, Germany}

\ead{goette@pks.mpg.de}

\begin{abstract}
The law of reflection and Snell's law are among the tenets of geometrical optics. 
Corrections to these laws in wave optics are respectively known as the angular Goos-H\"anchen shift and Fresnel filtering. 
In this paper we give a positive answer to the question of whether the two effects are common in nature and we study both effects in the more general context of optical beam shifts. 
We find that both effects are caused by the same principle, but have been defined differently.
We identify and discuss the similarities and differences that arise from the different definitions.
\end{abstract}


\vspace{2pc}
\submitto{\JO}
\maketitle

%
%

\section{Introduction}

It is not uncommon in physics that the same effect is discovered independently by different people and, as a consequence, may be known under two or more different names owing to differences in language and scientific tradition. 
A prominent and relevant example for this article is Snell's law, which in French speaking countries is more commonly know as Descartes's law, to remind us that it was Ren\'e Descartes who first published it in modern times \cite{Descartes:IM:1637} (although Ibn Sahl, Willebrord Snellius and others derived the result before him \cite{Rashed:Isis81:1990}). 

\begin{figure}
\centering
  \includegraphics[width=0.8\textwidth]{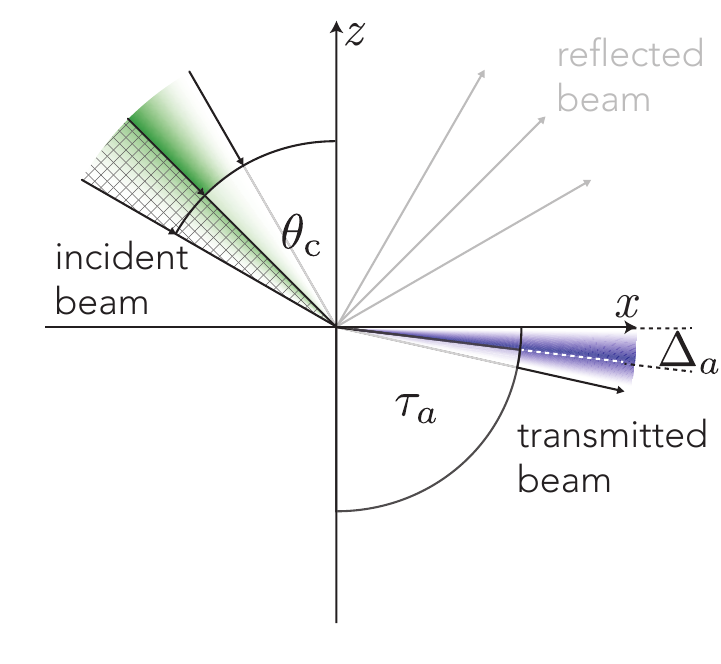}
  \caption{\label{fig:FresnelFiltering}%
  (Colour online) Schematic of Fresnel filtering at critical incidence.
  A light beam is incident from the region of higher refractive.
  The central wave vector impinges on the interface precisely at the critical angle of total internal reflection.
  All plane waves of the incident angular spectrum (green gradient) with larger angles of incidence (crossed) are fully reflected, whereas plane waves with smaller angles of incidence are transmitted (blue gradient).
  The gradient indicates the amplitude of the angular spectrum.
  Right at the critical angle the transmittance is zero which is why the transmitted beam has the maximum shifted away from grazing transmission.
   For clarity the reflected beam is shown as arrows only.   %
  }
\end{figure}

The deviation from the predictions of Snell's law in wave optics is known as `Fresnel filtering' (FF) \cite{TureciStone:OL27:2002} and describes a correction to the angle of refraction, which is particularly prominent for focussed beams incident close to the critical angle of total internal reflection. 
The origin of this effect is easily understood from the following simple picture (see Fig \ref{fig:FresnelFiltering}): assuming a symmetric angular spectrum and that the central wave vector of the focussed beam be incident exactly at the critical angle, all plane waves of the incident beam with an angle of incidence larger than the critical angle are fully reflected.
On the other hand, all plane waves incident below critical incidence are transmitted.
The magnitude of the effect is then the difference between the angle of refraction expected from geometrical or ray optics and the far field angle of the transmitted field.
If the incident angular spectrum is large, so is the transmitted spectrum, which explains intuitively why the effect of Fresnel filtering depends on the angular width of the incident beam.

\begin{figure}
\centering
  \includegraphics[width=0.8\textwidth]{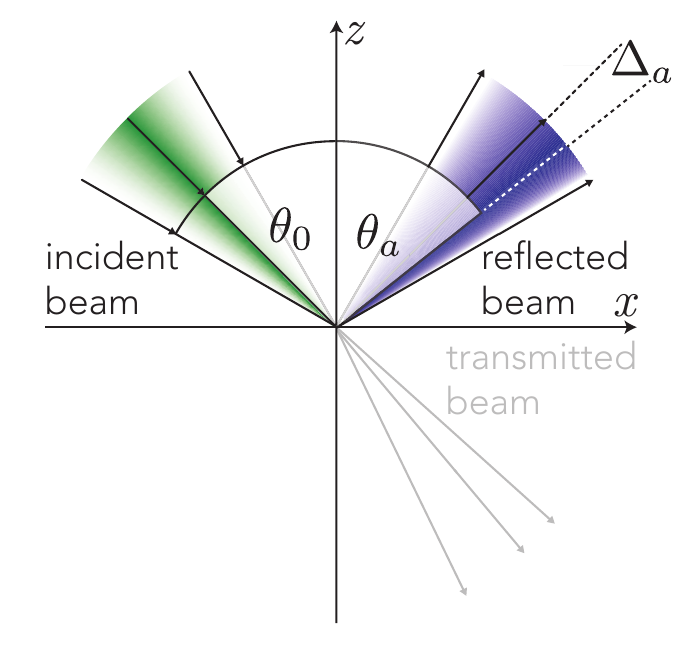}
  \caption{\label{fig:AngularGH}%
   (Colour online) Schematic of the angular Goos-H\"anchen (GH) shift.
  A lightbeam is reflected from a region of higher refractive index.
  Upon reflection each plane wave acquires a real weighting by virtue of the angle-dependent reflection coefficient.
  This changes the angular spectrum and as a consequence the mean angle of the reflected beam is different from the angle of incidence.
  }
\end{figure}

Independently from Fresnel filtering a family of optical beam shift effects has been studied mainly on reflection \cite{BliokhAiello:JO:2012}. 
In contrast to Fresnel filtering, which has been discovered in the context of two dimensional, scalar fields in optical microcavities \cite{
Rex+:PRL88:2002,Shinohara+:OL36:2011}, optical beam shifts have been explored in all three dimensions. 
Confined within the plane of incidence and therefore comparable to the 2D fields in microcavities, are the spatial and angular Goos-H\"anchen shift (GH) \cite{GoosHaenchen:AndP436:1947, Merano+:NatPhot3:2009} and it is the latter which concerns us here as a related effect to FF.
In its simplest form the angular GH shift pertains to generic incidence angles, that is not for the critical angle or the Brewster angle, although these special cases have been studied too \cite{ChanTamir:OL10:1985,AielloWoerdman:arxiv:2009,Berry:PRSLA467:2011}.
Using the same simple picture as above, but for generic incidence (see Fig. \ref{fig:AngularGH}), the angular GH shift can be explained as a weighting in the Fourier spectrum of the incident beam upon reflection, which leads to a shift of the mean angle in the reflected beam.
Unsurprisingly, the magnitude of the angular GH shift also depends on the angular spectrum.

One difference between the two effects is based on the different definitions as peak and mean angle of the transmitted field.
Another distinction between FF and the angular GH shift is rooted in their importance for the different scientific communities.
For example, the presence of regular, triangular-type resonator modes in chaotic microcavities of spiral shape could only be explained by the FF effect at near-critical reflection \cite{Altmann+:EPL84:2008}.
Such quasi-scarred resonances \cite{Lee+:PRL93:2004,Kwon+:OL31:2006} had been observed in wave calculations and, implementing the FF effect into an extended ray dynamics, could now be understood in terms of ray optics.
(Notice that implementing the FF effect alters the ray optics from Hamiltonian to become non-Hamiltonian \cite{Altmann+:EPL84:2008,UnterhinninghofenWiersig:PRE82:2010}, an interesting feature that is, however, beyond the scope of the present paper).
This example of FF in quasi-scarred modes illustrates nicely two properties of the effect:
(i) The FF effect is noticeable  especially around the critical angle, which, given a refractive index of $n=2$ inside the cavity, happened to be the inner angle of the triangular-type quasi-scarred modes.
(ii) The FF effect depends on the resonance wave number, implying a slight rotation between adjacent resonances in order to adjust the resonance geometry to the slight change in the FF correction.

In beam optics, on the other hand, the angular shift had been found to occur in the same setting as the spatial GH shift, but for incidence below the critical angle \cite{Ra+:SIAP24:1973,AntarBoerner:AP7:1975}.
As the angular GH shift is a deflection of the beam's propagation direction the displacement of the beam caused by this effect scales with the distance from the interface.
This is why in situations where both spatial and angular GH shifts occur, for example in metallic reflection \cite{Goette+:OE16:2008b}, the contribution of the angular shift to the overall displacement will be become dominant when looking sufficiently far down the beam \cite{Aiello+:PRA80:2009}.
Recently, it has been found that for higher order Laguerre-Gaussian beams, or more generally any vortex beam, the total spatial shift can be explained by a mixing of the spatial and angular shifts of the fundamental beam \cite{Bliokh+:OL34:2009,Merano+:PRA82:2010,DennisGoette:NJP14:2012}.

FF and the angular GH shift are both counterparts to \emph{spatial} GH shift.
Either as two independent directions in phase phase \cite{SchomerusHentschel:PRL96:2006}, or as real and imaginary part of the same complex quantity namely the logarithmic derivative of the reflection coefficients \cite{Artmann:AndP437:1948,ChanTamir:OL10:1985,Merano+:NatPhot3:2009,GoetteDennis:NJP14:2012}. 
As such the formulas for the angular and spatial GH shift on reflection are
\begin{equation}
\label{eq:rgh}
\Delta^\mathrm{GH}_a = \sigma_2 \mathrm{Re}  \left. \frac{r'}{r} \right|_{\theta_0}  \quad \mbox{and} \quad D^\mathrm{GH}_a  = -\frac{1}{k} \mathrm{Im}  \left. \frac{r'}{r} \right|_{\theta_0},
\end{equation}
where $r$ is the reflection coefficient \cite{Jackson:JohnWiley:1998,BornWolf:CUP:2003} and the prime denotes the derivative with respect to the incident angle at the central angle $\theta_0$. These shifts have their natural units, given by the angular width or variance $\sigma_2$ for the angular shift and the inverse of the wavenumber $k$ for the spatial shift.

In this paper we focus on the differences and similarities between the angular Goos-H\"anchen (GH) shift and Fresnel filtering.
The arrangement of the paper follows our logical outline of the argument.
In the first part of the next section we derive an explicit expression for the angular GH shift on transmission for generic incidence which we compare to the known formula for the corresponding shift on reflection.
In the second part we find an explicit expression for critical incidence and we compare it with the known results from FF at critical incidence.
Section \ref{sec:FF} contains derivations of analytical formulas for FF at a generic angle of  incidence, which we use for a concluding comparison in a discussion in section \ref{sec:discussion}.

%
%

\section{Angular GH on transmission}
\label{sec:angularGH}

Our derivation for the expression of the angular GH shift in transmission relies on the concept of the virtual beam as introduced in \cite{GoetteDennis:NJP14:2012}.
However, whereas for the case of reflection the virtual beam is obtained by a specular reflection of every plane wave, that is the reflection coefficient is $r = \pm 1$ throughout, for transmission the virtual refracted beam is obtained by applying Snell's law for every plane wave according to the local angle of incidence.
The spatial and angular shifts on transmission are then the differences to the virtual reflected beam due to the inclusion of the appropriate transmission coefficient.

\begin{figure}
\centering
  \includegraphics[width=0.8\textwidth]{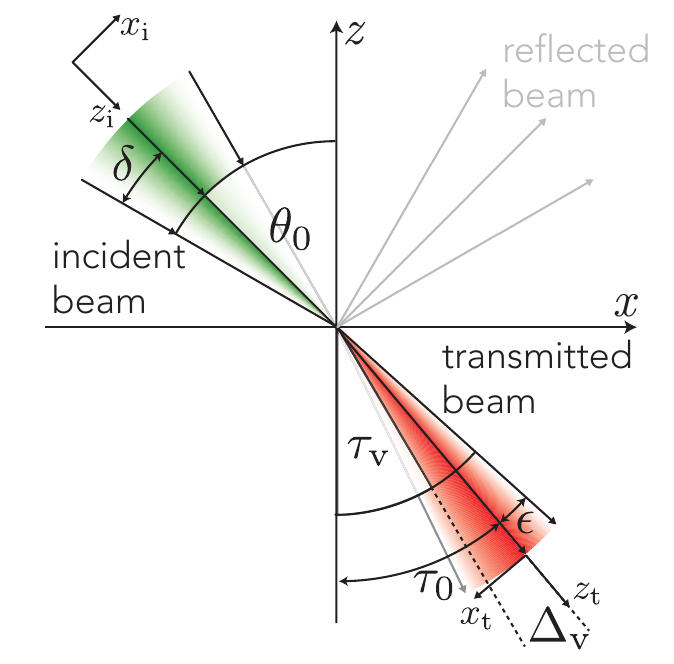}
  \caption{\label{fig:virtual}%
  (Colour online) Construction of the virtual transmitted beam. 
   Every incident plane wave is refracted according to Snell's law, which introduces an asymmetry in the transmitted spectrum.
   Averaged over the whole spectrum this asymmetry results in an angular shift even for the virtual transmitted beam.
   The figure also shows the coordinate system and angles used in section \ref{sec:FF}.
   }
\end{figure}

In 2D we can treat the polarisation orthogonal to plane of incidence (s  or TM) and parallel to it  (p  or TE) separately,\footnote{As there is always a confusion between the different labels for polarisation, we state explicitly that p (TE) is the polarisation for which there exists a Brewster angle in partial reflection. Our identification of s being TM and p being TE follows common usage in the microcavity community where the reference plane is the 2D resonator and not the plane of incidence.} and the incident and virtual transmitted beam are the same for both polarisations \cite{Jackson:JohnWiley:1998}; the only place where the polarisation enters is the choice of the appropriate transmission coefficient.
We adopt the notation of a reduced refractive index $n = n_2/n_1$, where $n_1$ is the index of the medium in which the incident beam and reflected beam propagate.
Using $x$ for the spatial coordinate along the interface and $z$ for the coordinate normal to it, we parametrize the incident beam by means of the angle of incidence $\theta$ for each plane within the spectrum $\sigma(\theta)$
\begin{equation}
  \psi_\mathrm{i} = \int d\theta \sigma(\theta) \mathrm{e}^{\mathrm{i} k \left( x \sin \theta - z \cos \theta \right)}.
\end{equation}
The virtual transmitted beam is constructed by changing the direction of every plane wave in accordance with Snell's law $\sin \theta = n \sin \tau$ while maintaining the parametrization of the beam by $\theta$ (see Fig. \ref{fig:virtual}):
\begin{equation}
  \psi_\mathrm{v} = \int  d\theta \sigma(\theta) \mathrm{e}^{\mathrm{i} k \left(x \sin \theta - z \sqrt{n^2 - \sin^2 \theta}\right)}.
\end{equation}
If the incident beam is in the optically thicker medium, the virtual beam includes evanescent waves.
For the calculation of the angular GH shift, however,  we only include propagating plane waves, which can be enforced by setting the upper bound of the integration to the critical angle $\theta_\mathrm{c}$.
The real transmitted beam is obtained from the virtual beam by inserting the appropriate transmission coefficients for s (TM) and p (TE) polarisation \cite{BornWolf:CUP:2003}
\begin{eqnarray}
  \label{eq:tscoefficients}
  t_\mathrm{s} & = \frac{2 \cos \theta}{\cos \theta + \sqrt{n^2 - \sin^2 \theta}} \quad \mbox{(TM)},\\
  \label{eq:tpcoefficients}
  t_\mathrm{p} & = \frac{2 n \cos \theta}{n^2 \cos \theta + \sqrt{n^2 - \sin^2 \theta}} \quad \mbox{(TE)},
\end{eqnarray}
which yields
\begin{equation}
  \psi_a = \int  d\theta \sigma(\theta) t_a (\theta) \mathrm{e}^{\mathrm{i} k \left(x \sin \theta - z \sqrt{n^2 - \sin^2 \theta}\right)},
\end{equation}
where $a = \mathrm{s (TM)}, \mathrm{p (TE)}$. 

\begin{figure}
\centering
  \includegraphics[width=0.8\textwidth]{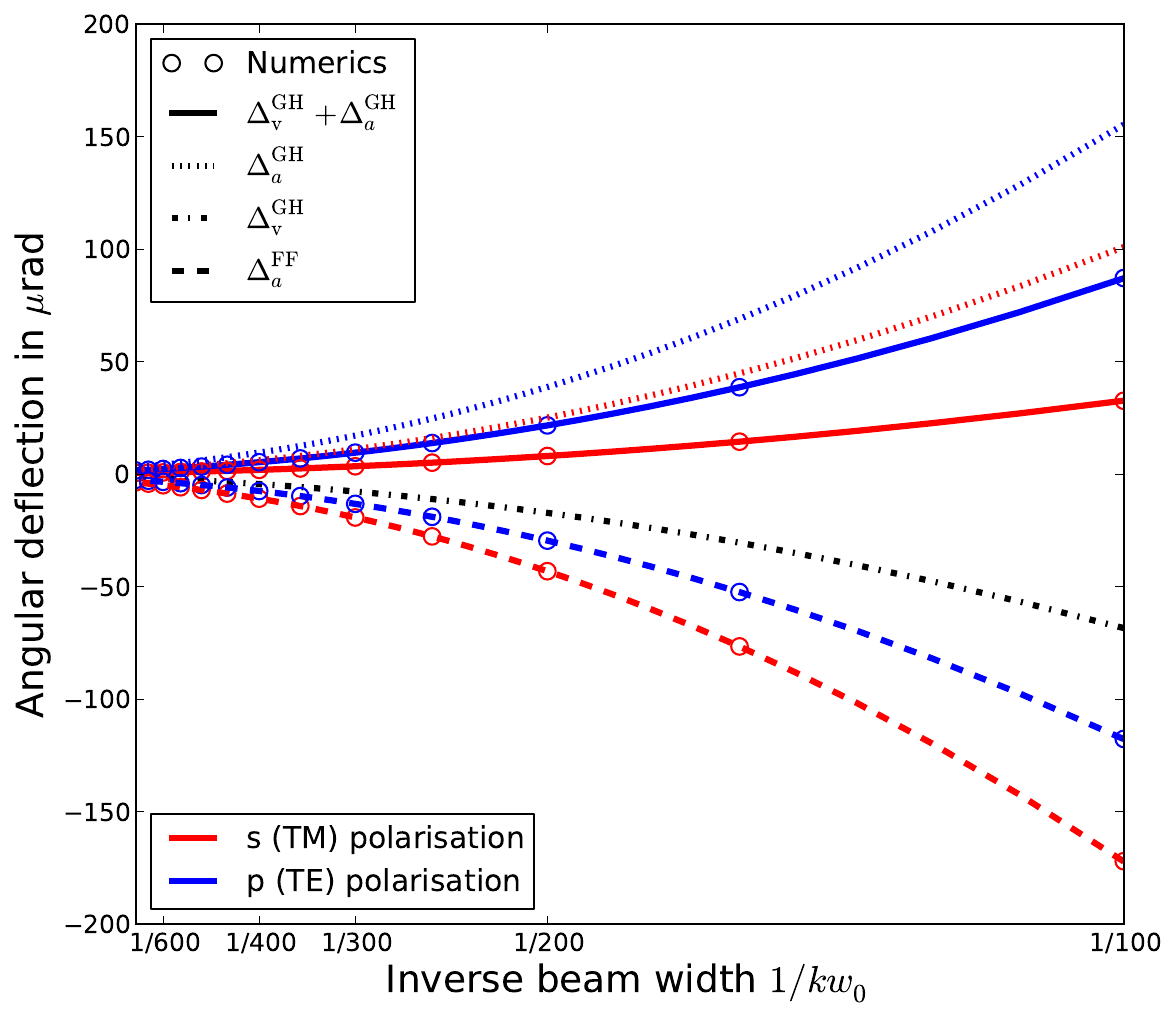}
  \caption{\label{fig:GenericAGH}%
  (Colour online) Angular GH shift at a generic angle of incidence.
  The plot compares the angular GH shift with FF for both s (red) and p (blue) polarisation at an incidence angle $\theta_0 = 0.4$rad ($\approx 23^\circ$) and with $n = 2/3$.
  The GH shifts are broken down into the total shift $\Delta_\mathrm{v}^\mathrm{GH} + \Delta_a^\mathrm{GH}$ (solid), the virtual shift $\Delta_\mathrm{v}^\mathrm{GH}$ (dashed-dotted) and the angular GH shift $\Delta_a^\mathrm{GH}$ (dotted).
  The formulas for the FF shifts $\Delta_a^\mathrm{FF}$ (dashed) are derived in section \ref{sec:FF}.
   }
\end{figure}

Following the treatment in \cite{GoetteDennis:NJP14:2012} we calculate the mean angle of the transmitted beam $\psi_\mathrm{t}$ as centroid of the transmitted angular spectrum.
Unlike for reflection, however, we can observe a deviation from the laws of geometrical optics even for the virtual beam.
Because the incident and transmitted beam are in two different media, the intensity is no longer simply proportional to the modulus of the amplitude \cite{BornWolf:CUP:2003}.
To account for this distinction we introduce a virtual transmittance $\mathcal{V} = n | \cos \tau / \cos \theta |$, where $\tau \equiv \sin^{-1} (\sin \theta / n)$ denotes the angle of transmitted plane waves.
As we only consider propagating waves in transmission, all the angles and the transmission coefficients are real, so that we can ignore the moduli.
This also implies that there will be no \emph{spatial} GH shift in this setting.
The mean angle for the virtual transmitted beam is thus calculated as centroid of the virtual transmitted spectrum:
\begin{equation}
\label{eq:meanvirtual}
  \tau_\mathrm{v} = \langle \tau (\theta) \rangle_\mathrm{v} = \frac{\int d\theta \: |\sigma(\theta)|^2 \mathcal{V}(\theta) \, \tau (\theta)}{\int d\theta \: |\sigma(\theta)|^2 \mathcal{V}(\theta)}.
\end{equation}
Assuming that the spectrum $\sigma$ of the incident beam is symmetric and narrowly concentrated around the central plane wave with incidence $\theta_0$ we expand all $\theta$ dependent quantities by setting $\theta = \theta_0 + \delta$ with $\delta$ being small.
Identifying $\tau (\theta_0) \equiv \tau_0$ we write explicitly
\begin{eqnarray}
  \mathcal{V} (\theta_0+ \delta) & \approx  n \frac{\cos \tau_0}{\cos \theta_0} \left[ 1 + \delta \left( \tan \theta_0 - \tau' \vert_{\theta_0} \tan \tau_0 \right) + \dots \right] , \\
  \label{eq:refraction}
  \tau (\theta_0+ \delta) & \approx \tau_0 + \delta \, \tau' \vert_{\theta_0} + \frac{\delta^2}{2} \tau'' \vert_{\theta_0} + \dots,
\end{eqnarray} 
where the prime denotes derivatives with respect to $\theta$.
On substitution of these expansions into (\ref{eq:meanvirtual}) we can identify the zeroth order in $\delta$ with Snell's law for the central plane wave, while the first correction term is of order $\delta^2$ as the first order terms vanish on integration due to symmetry.
Collecting all terms of order $\delta^2$ yields an approximate expression for the mean angle of the virtual beam
\begin{equation}
  \label{eq:tauvirtual}
  \tau_\mathrm{v} \approx \tau_0 + \left[ \tau' \vert_{\theta_0} ( \tan \theta_0 - \tau' \vert_{\theta_0} \tan \tau_0 )+ \frac{1}{2} \tau'' \vert_{\theta_0} \right] \langle \delta^2 \rangle =  \tau_0  -\frac{1}{2} \tau'' \vert_{\theta_0}   \langle \delta^2 \rangle,
\end{equation}
where the equality follows from the properties of $\tau(\theta)$ and $\langle \delta^2 \rangle$ is the second spectral moment or angular variance of the incident beam calculated as
\begin{equation}
  \langle \delta^2 \rangle = \frac{\int d\delta \: |\sigma(\theta_0 + \delta)|^2 \delta^2}{\int d\delta \: |\sigma(\theta_0 + \delta)|^2}.
\end{equation}
The second order derivative in (\ref{eq:tauvirtual}) is not present in other derivations of the angular Goos-H\"anchen shift on transmission (and it does not occur for reflection \cite{GoetteDennis:NJP14:2012,DennisGoette:NJP14:2012}). 
Our approach thus differs from known formulas \cite{BliokhAiello:JO:2012,Bliokh+:OL34:2009}. 
We support our results by the excellent agreement with a direct numerical calculation of the virtual shift as defined in (\ref{eq:meanvirtual}) (see Fig.~\ref{fig:GenericAGH}).
The virtual angular GH shift is thus given by
\begin{equation}
\label{eq:vgh}
\Delta^\mathrm{GH}_\mathrm{v} = \tau_\mathrm{v} - \tau_0.
\end{equation}
This deviation from Snell's law stems from the asymmetric refraction of the plane waves: whereas the incident spectrum is symmetrically distributed around the central plane wave $\theta_0$, refraction introduces an asymmetry which leads to a shift or deflection of the mean angle even for the virtual transmitted beam.

For the real transmitted beam we have to exchange the virtual transmittance $\mathcal{V}$ with the real transmittance $\mathcal{T}_a, a = \mathrm{s (TM)},\mathrm{p (TE)}$, which is defined as \cite{Jackson:JohnWiley:1998}:
\begin{equation}
  \mathcal{T}_a(\theta) =  n \left|\frac{\cos \tau}{\cos \theta} \right| |t_a (\theta) |^2
\end{equation}
and includes the transmission coefficients.
The mean angle of the real reflected beam for either s or p polarisation is the centroid of the real transmitted spectrum:
\begin{equation}
\label{eq:meanreal}
  \tau_a =  \langle \tau(\theta) \rangle_a = \frac{\int d\theta \: |\sigma(\theta)|^2 \mathcal{T}_a(\theta) \, \tau (\theta)}{\int d\theta \: |\sigma(\theta)|^2 \mathcal{T}_a(\theta)} \quad a = \mathrm{s (TM)}, \mathrm{p (TE)}.
\end{equation}
In this case we expand the transmittance $\mathcal{T}_a, a = \mathrm{s},\mathrm{p}$ around $\theta_0$
\begin{equation}
  \label{eq:transmittance}
  \mathcal{T}_a (\theta_0 + \delta) \approx  n \frac{\cos \tau_0}{\cos \theta_0} t_a^2 \left[1 + \delta \left( 2 \frac{t_a'}{t_a}  \bigg\vert_{\theta_0} + \tan \theta_0 - \delta \, \tau' \vert_{\theta_0} \tan \tau_0 \right) + \dots \right],
\end{equation} 
where the prime denotes derivatives with respect to $\theta$. 
The mean angle of the transmitted beam is therefore approximately
\begin{equation}
\label{eq:realshift}
  \tau_a \approx \tau_0 + \left[ 2 \tau' \vert_{\theta_0} \frac{t_a'}{t_a} \bigg\vert_{\theta_0} - \frac{1}{2} \tau'' \vert_{\theta_0} \right] \langle \delta^2 \rangle \quad a = \mathrm{s (TM)}, \mathrm{p (TE)}.
\end{equation}   
The angular GH shift is the difference between the real and virtual transmitted angle $\Delta^\mathrm{GH}_a = \tau_a - \tau_\mathrm{v}$ for $a = \mathrm{s}, \mathrm{p}$, which yields
\begin{equation}
\label{eq:agh}
  \Delta^\mathrm{GH}_a(\theta_0) = 2 \, \tau' \vert_{\theta_0}  \frac{t_a'}{t_a} \bigg\vert_{\theta_0} \langle \delta^2 \rangle  =  \frac{2 \cos \theta}{\sqrt{n^2 - \sin^2 \theta}}  \frac{t_a'}{t_a} \bigg\vert_{\theta_0} \langle \delta^2 \rangle \quad a=\mathrm{s},\mathrm{p}.
\end{equation}
This formula is the main result of this section. In this form the angular GH on transmission is analogous to its counterpart on reflection (\ref{eq:rgh}) apart from the geometrical factor $\cos \theta_0/\cos \tau_0$. In particular it depends linearly on the second moment of the angular spectrum.
We also define the total angular GH shift as the sum of (\ref{eq:vgh}) and (\ref{eq:agh}) as
\begin{equation}
\label{eq:totalgh}
  \Delta^\mathrm{GH}_\mathrm{v}+ \Delta^\mathrm{GH}_a =  \left( \tau' \vert_{\theta_0} \frac{t_a'}{t_a} \bigg\vert_{\theta_0}  -\frac{1}{2} \tau'' \vert_{\theta_0}  \right) \langle \delta^2 \rangle  \quad a=\mathrm{s},\mathrm{p},
\end{equation}
which will be compared to the FF in section \ref{sec:FF}.

Of course, the distinction between the angular deflection for the virtual beam and the real transmitted beam is to some degree artificial and a measurement of the mean angle would give a result corresponding to (\ref{eq:totalgh}).
However, actual measurements are often differential, say for different polarisations \cite{Merano+:NatPhot3:2009}.
In this case the virtual shift cancels and the measured quantity corresponds to $\Delta^\mathrm{GH}_s -  \Delta^\mathrm{GH}_p$.
For the purpose of this paper it is useful to identify the constituent parts of the angular GH shift.
To this end we compare in Fig.~\ref{fig:GenericAGH}  the total angular GH shift $\Delta^\mathrm{GH}_\mathrm{v} + \Delta^\mathrm{GH}_a$ for both polarisations $a=\mathrm{s}, \mathrm{p}$ with numerical calculations based on (\ref{eq:meanreal}) for a generic angle of incidence ($\theta_0 = 0.4$rad or $\approx 23^\circ$) of a Gaussian beam with a width $w_0$  at a glass/air interface ($n=2/3$).
We also show the angular GH shift as defined in (\ref{eq:agh}) and the virtual shift alone.
From this plot it is obvious that the virtual and real angular GH shift are competing effects, which leads to cancellations for the total shift.

In \cite{TureciStone:OL27:2002} the magnitude of the FF at critical incidence is given explicitly in terms of the beam waist.
In contrast to generic incidence, for the critical angle the effect is not inversely proportional to the square of the beam waist, but only to the square root of it.
It is interesting to see whether the angular GH shift reproduces this result.
However, because right at the critical angle both the transmission coefficients (\ref{eq:tscoefficients}, \ref{eq:tpcoefficients}) and the angle of refraction (\ref{eq:refraction}) change from real to complex functions, the expansions (\ref{eq:refraction}) and (\ref{eq:transmittance}) change in nature.
On setting $\theta = \theta_\mathrm{c} - \delta$ and restricting ourselves to positive angles below the critical angle, we can expand the refracted angle in fractional powers of $\delta$ \cite{Cutkosky:AMS:2004}:
\begin{equation}
\label{eq:taucritical}
  \tau(\theta_\mathrm{c}) = \frac{\pi}{2} - \sqrt{\frac{2}{\tan \theta_\mathrm{c}}} \sqrt{\delta} + O[(\sqrt{\delta})^3].
\end{equation}
The transmittance can also be expanded in fractional powers of $\delta$ at the critical angle, though because $\mathcal{T}_a$ is present in both the numerator and denominator of (\ref{eq:realshift}), and because we consider only the first correction term, the coefficients cancel.

Within the integral (\ref{eq:realshift}) the zeroth order of (\ref{eq:taucritical}) gives rise to Snell's law at critical incidence, namely an angle of refraction of $\pi/2$.
The first correction term is essentially given by the term of order $\sqrt{\delta}$ in the expansion (\ref{eq:taucritical}), but picks up another $\sqrt{\delta}$ from the expansion of the transmittance.
The mean angle of the transmitted beam is thus given by:
\begin{equation}
\label{eq:criticalintegral}
  \tau_a \approx \frac{\pi}{2} - \sqrt{\frac{2}{\tan \theta_\mathrm{c}}} \frac{\int_0^\infty d\theta \: |\sigma(\theta_\mathrm{c} - \delta)|^2  \, \delta}{\int_0^\infty d\theta \: |\sigma(\theta_\mathrm{c} - \delta)|^2 \sqrt{\delta}}  \quad a = \mathrm{s (TM)}, \mathrm{p (TE)},
\end{equation}
which holds for both s (TM) and p (TE) polarisation. 
For a general spectrum a meaningful interpretation in terms of moments of the spectrum is no longer possible in this form, but for a Gaussian beam with spectrum $\sigma(\delta) = \exp[-(k w_0)^2 (\theta_\mathrm{c} - \delta)^2/4]$, the integrals can be calculated.
The angular GH shift for transmission at critical incidence is given by
\begin{equation}
\label{eq:criticalAGH}
 (\Delta^\mathrm{GH}_\mathrm{v}+\Delta^\mathrm{GH}_a) \big\vert_{\theta_\mathrm{c}} = - \sqrt{\frac{2}{\tan \theta_\mathrm{c}}} \frac{2^{1/4}}{\Gamma(3/4)} \frac{1}{\sqrt{k w_0}},
\end{equation}
where $\Gamma$ is the Gamma function \cite{DMLF}. 

Comparing the result with Tureci and Stone \cite{TureciStone:OL27:2002} (see section \ref{sec:FF}) we find the same dependence on the square root of the width of the beam, but the proportionality factor is different.
However, the difference is fairly small, as the factor $\frac{2^{1/4}}{\Gamma(3/4)} \approx 0.97$ and therefore close to unity.
In their original paper Tureci and Stone state a FF effect of around $30^\circ$ at the critical incidence for a refractive index of $n=1/1.56$ and a scaled beam width of $k w_0 = 8.82$.  
Evaluating (\ref{eq:criticalAGH}) for these parameter values gives an angular deflection of about $29^\circ$ which is in line with the different factors.

This observation is interesting as the difference in the definition of FF and angular GH as peak and mean angle is of particular importance at critical incidence as only half of the incident spectrum is transmitted.
The transmitted spectrum is therefore no longer Gaussian, which would enhance the difference between peak and mean angle.
A fuller account of the shifts at the critical angle which could explain this behaviour is in preparation.

\begin{figure}
\centering
  \includegraphics[width=0.8\textwidth]{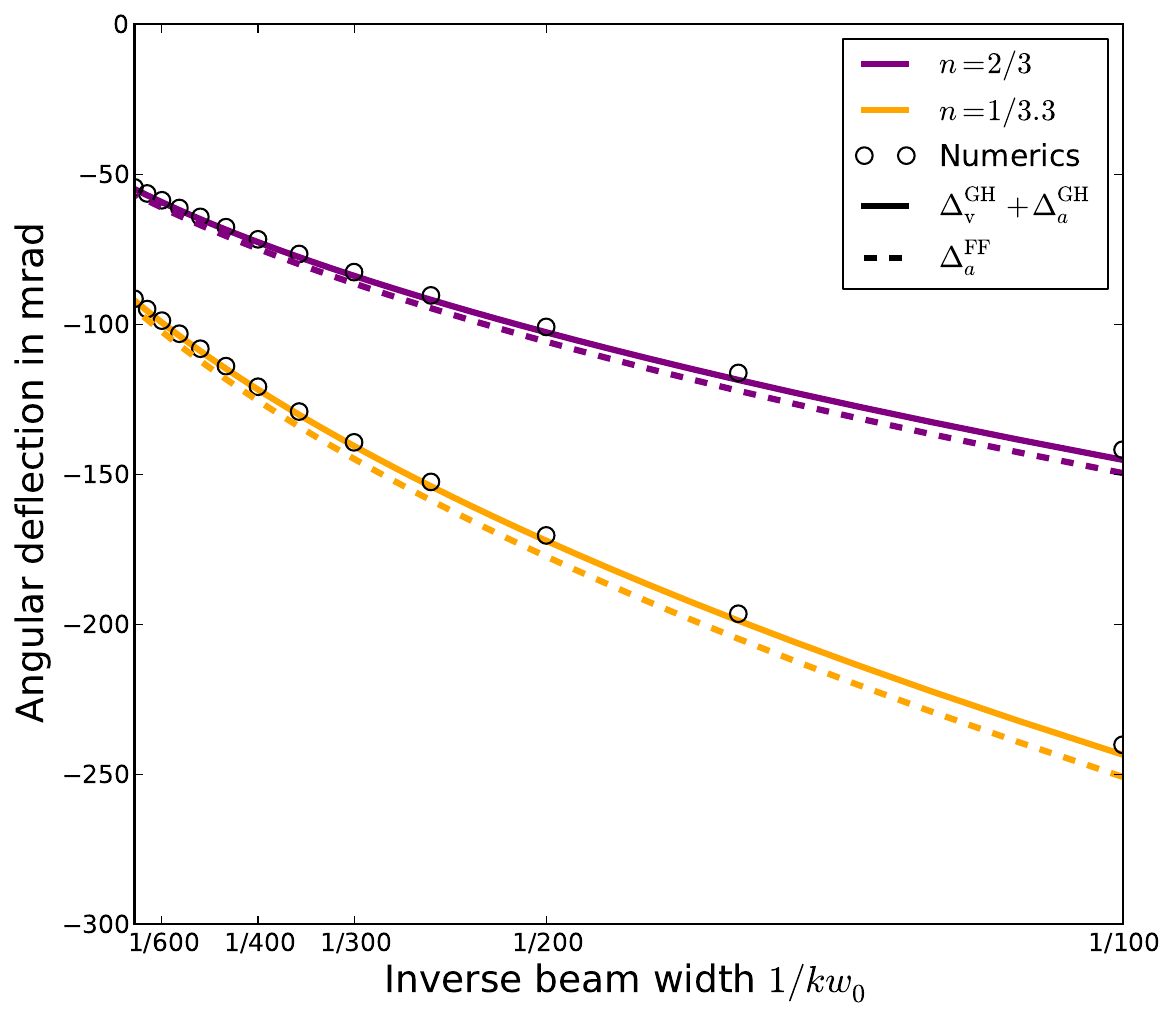}
  \caption{\label{fig:CriticalAGH}%
  (Colour online) Angular GH shift at a critical incidence.
  The plot compares the angular GH shift with FF at critical incidence $\theta_\mathrm{c}= \arcsin(n)$ for two different refractive interfaces: glass - air with $n = 2/3$ (purple) and GaAs - air with $n = 1/3.3$ (orange).
  The shift is identical for both s and p polarisation.
  The plot compares the total angular GH shift $\Delta_\mathrm{v}^\mathrm{GH} + \Delta_a^\mathrm{GH}$ (solid) with the shift due to FF $\Delta_a^\mathrm{FF}$ (dashed).
  At critical incidence the shift is 3 orders of magnitude larger compared with the shifts in Fig.~\ref{fig:GenericAGH}.
   }
\end{figure}

In Fig.~\ref{fig:CriticalAGH} we show a comparison between our formula (\ref{eq:criticalAGH}) and a numerical calculation of (\ref{eq:criticalintegral}) for two different refractive indices.
The agreement is slightly worse than for general incidence, but nevertheless excellent.
The fact that the shift is negative and does not depend on the incident polarisation suggest that the angular GH shift at critical incidence is predominantly given by the virtual shift.

%
%

\section{Fresnel filtering}
\label{sec:FF}

In this section, we study the beam shift defined for the field and not for the intensity as in the previous section.
The deviation from Snell's law for the field amplitude has been studied by Tureci and Stone, who named this effect `Fresnel filtering' (FF) \cite{TureciStone:OL27:2002}.
To make this paper self-contained, we briefly review their theory.
We consider a situation where a Gaussian beam with central incident angle $\theta_0$ is propagating from a thicker medium with refractive index $n_1$ to a medium with refractive index $1$, such that $n=1/n_1$, and scattered by an infinite planar interface dividing the two media.
We adopt the coordinate systems $(x_\mathrm{i}, z_\mathrm{i})$ and $(x_\mathrm{t}, z_\mathrm{t})$ attached
to the incident and refracted beams, as shown in Fig.~\ref{fig:virtual}.
Within the coordinate system $(x_\mathrm{i}, z_\mathrm{i})$, the incident Gaussian beam with a beam waist $w_0$ is given by
\begin{equation}
\psi_\mathrm{i}(x_\mathrm{i}, z_\mathrm{i}) =\frac{E_0 w_0}{w(z_\mathrm{i})}\exp\left[ -\left(
\frac{x}{w(z_\mathrm{i})} \right)^2+\mathrm{i} k z_\mathrm{i} \right],
\end{equation}
where $w^2(z_\mathrm{i})=w_0^2-\mathrm{i}\frac{2z_\mathrm{i}}{k}.$
We assume that the position of the beam waist is located at the interface.

By virtue of the angular spectrum representation, this beam can be expressed as a superposition of plane waves
\begin{equation}
  \psi_\mathrm{i}(x_\mathrm{i}, z_\mathrm{i}) = \frac{k w_0 E_0}{2\sqrt{\pi}} \int ds\,\sigma(s)\exp\left[ \mathrm{i} k\left( x_\mathrm{i} \sin \delta + z_\mathrm{i} \cos \delta \right) \right],
\end{equation}
where $\delta$ depends on $s$ via $\sin\delta=s$ and  $\cos\delta =\sqrt{1-s^2}.$ 
The angular spectrum $\sigma$ is given by $\sigma(s)=\exp[ -(kw_0 s/2)^2]$.
The transmitted beam is then given by
\begin{equation}
  \psi_a (x_\mathrm{t}, z_\mathrm{t})  = \frac{ k w_0 E_0}{2\sqrt{\pi}}\int ds\, t_a (s)\, \sigma(s)\exp\left[ \mathrm{i} n k\left( x_\mathrm{t} \sin\epsilon +z_\mathrm{t} \cos\epsilon \right) \right],
\label{eq:E_e}
\end{equation}
for $a=\mathrm{s (TM)},\mathrm{p (TE)}$ polarisation.
Here, $\delta$ and $\epsilon$ are related through $\sin(\theta_0+\delta) = n \sin(\tau_0+\epsilon)$, which also defines $\epsilon$ as a function of $s$, and $t_a(s)$ is the appropriate Fresnel transmission coefficient (\ref{eq:tscoefficients}, \ref{eq:tpcoefficients}) with $\theta(s) = \theta_0 + \delta(s)$.
On using the polar coodinates $x_\mathrm{t} = \rho \sin(\tau - \tau_0)$ and $z_\mathrm{t} = \rho \cos(\tau - \tau_0)$ the refracted beam 
(\ref{eq:E_e}) can be rewritten as
\begin{equation}
  \psi_a (\rho, \tau) = \frac{k w_0 E_0}{2\sqrt{\pi}}\int ds\, t_a(s)\,\sigma(s)\exp\left[\mathrm{i} k\rho\cos(\tau - \tau_0 - \epsilon))\right],
\end{equation}
where $\epsilon$ depends on $s$ via Snell's law.
In the far field, that is in the limit $\rho\to\infty$, the integrand is highly oscillating and can be evaluated using the saddle-point method. As pointed out by Tureci and Stone the relevant saddle point is where the cosine in the integrand takes its maximum. 
This relates the free variable $\tau$ to $\tau_0 + \epsilon$ and via Snell's law also to $\theta_0 + \delta$.
For a given $\tau$ the saddle point $s_0$ is thus determined by solving $\sin[\theta_0+\delta(s_0)]=n \sin\tau$ for $s_0.$
Upon expansion of the exponent and subsequent integration the transmitted field is given by
\begin{equation}
  \psi_a(\rho,\tau) \approx \frac{kw_0 E_0}{\sqrt{2\mathrm{i} k \rho}}\, t_a(s_0) \sigma(s_0) J(\tau; s_0) \mathrm{e}^{\mathrm{i} k\rho}
\label{eq:tGJ}
\end{equation}
in the far field. The term $J(\tau; s_0)$ is defined as
\begin{equation}
  J(\tau; s_0) = n \frac{\cos\tau}{\sqrt{1 - n^2 \sin^2 \tau}} \sqrt{1-s_0^2} =  n \frac{\cos\tau}{\cos \theta} \sqrt{1-s_0^2}.
\end{equation}
Eq.~(\ref{eq:tGJ}) describes the transmitted field as a function of $\tau$ and for a saddle point $s_0$ which, for a given $\tau$, is fully determined by the incident angle $\theta_0$ and the refractive index $n$. 

From (\ref{eq:tGJ}), one can estimate the peak position of the refracted beam for certain cases.
First, we consider the case for the critical incidence (i.e. $\theta_0=\theta_c$), which has already been  treated by Tureci and Stone \cite{TureciStone:OL27:2002}.
We put $\tau=\pi/2-\epsilon$ with $\epsilon$ being a small parameter.
From $\sin[\theta_0+\delta(s_0)]=n \cos \epsilon$, we have
\begin{equation}
  s_0 \approx -\frac{n \epsilon^2}{2\sqrt{1 - n^2}}.
\end{equation}
Inserting this into $t_a(s_0)$, $\sigma (s_0)$, and $J(\tau; s_0)$, we get for both p (TE) and s (TM) polarisation the same proportionality in the leading order in $\epsilon$
\begin{equation}
  t_a(s_0) \sigma(s_0) J(\tau; s_0) \propto \epsilon \exp\left[ -\frac{n^2}{4(1 - n^2)} \left( \frac{kw_0}{2} \right)^2 \epsilon^4 \right].
\end{equation}
The different proportionality factor for p and s polarisation does not affect the maximum of this expression.
Finding the maximum determines the peak position of the electric field as $\pi/2 - \epsilon$; the magnitude of the FF is then given by the difference between this peak position and the angle expected from Snell's law:
\begin{equation}
  \Delta_a^\mathrm{FF} = -\left( \frac{1 - n^2}{n^2}\right)^{1/4}\sqrt{\frac{2}{kw_0}} = - \sqrt{\frac{2}{\tan\theta_c}}\frac{1}{\sqrt{kw_0}}.
\end{equation}
This is Tureci and Stone's original result and shows that the angular shift scales as $(kw_0)^{-1/2}$ at critical incidence. A comparison with the corresponding expression for the angular GH shift at critical incidence (\ref{eq:criticalAGH}) is shown in Fig.~\ref{fig:CriticalAGH}.

For generic incidence Tureci and Stone did not give an explicit analytic result which could be compared to the sum $\Delta_\mathrm{v}^\mathrm{GH} + \Delta_a^\mathrm{GH}$ [see (\ref{eq:vgh}) and (\ref{eq:agh})].
In the remainder of this section we therefore derive an expression for the magnitude of the Fresnel filtering for generic incidence.
Let us assume that the incident beam is narrow and the incident angle is far from the critical angle (i.e. $\delta \ll \theta_c - \theta_0$),
so that the profile of the refracted beam can be expected to be well approximated by a Gaussian distribution. Assuming that the refracted beam is also narrow, we put $\tau=\tau_0+\epsilon$ with $\epsilon$ being a small parameter.
Inserting this into $\sin[\theta_0+\delta(s_0)]=n \sin\tau$, we have $s_0 \approx (\cos\tau_0/\cos\theta_0) n\epsilon$.
Thus, the Gaussian distribution $\sigma (s_0)$ is written as
\begin{equation}
  \sigma(s_0) \approx \exp\left[ -\left(\frac{\cos\tau_0}{\cos\theta_0}\frac{kw_0}{2}\right)^2 n^2 \epsilon^2 \right].
\end{equation}
Now, let us consider how the functions $J(\tau; s_0)$ and $t_a(s_0)$ shift the above Gaussian distribution.
To do this, we expand $\ln J(\tau; s_0)$ and $\ln t_a(s_0)$ in terms of $\epsilon$ as follows:
\begin{eqnarray}
  \ln J & = & j^{(0)} + j^{(1)}\epsilon + j^{(2)} \epsilon^2 + \cdots, \\
  \ln t_a & = &  t^{(0)}_a+ t^{(1)}_a \epsilon + t^{(2)}_a \epsilon^2+\cdots.
\end{eqnarray}
Then, we have
\begin{equation}
  \fl \ln t_a(s_0) \sigma(s_0) J(\tau; s_0) = \left[ t^{2}_a+j^{(2)} - \left(\frac{\cos\tau_0}{\cos\theta_0} \frac{n kw_0}{2}\right)^2
\right] \epsilon^2 + (t^{(1)}_a + j^{(1)}) \epsilon + \cdots.
\end{equation}
Ignoring the terms higher than $\epsilon^2$ and completing the square on the right hand side, we can rewrite the above equation as
\begin{equation}
\ln t_a(s_0) \sigma(s_0) J(\tau;s_0) \approx C - \left(\frac{\cos\tau_0}{\cos\theta_0} \frac{n kw_0}{2}\right)^2 \left(\epsilon -  \Delta_a^\mathrm{FF}\right)^2,
\label{eq:ln_tGJ}
\end{equation}
where $C$ is a constant and for $kw_0\gg 1$. Here, $\Delta_a^\mathrm{FF}$ is given by
\begin{equation}
  \label{eq:FF}
  \Delta_a^\mathrm{FF} = 2 \left( \frac{\cos \theta_0}{\cos \tau_0} \right)^2  \left( \frac{1}{n k w_0} \right)^2 (t^{(1)}_a + j^{(1)}).
\end{equation}  
Eq. (\ref{eq:ln_tGJ}) tells us that the peak position of the refracted beam is shifted by $\Delta_a^\mathrm{FF}$ from the transmission angle $\tau_0$. 
The coefficients $t^{(1)}_a$ and $j^{(1)}$ can be written as
\begin{equation}
t^{(1)}_a = \left.\frac{t'_a}{t_a}\right|_{s(\epsilon=0)} = n \frac{\cos \tau_0}{\cos \theta_0} \left.\frac{t'_a}{t_a}\right|_{\delta=0},
\end{equation}
and
\begin{equation}
j^{(1)} = \left.\frac{J'}{J}\right|_{s(\epsilon=0)}=-
(1-n^2)\frac{\tan\tau_0}{\cos^2\theta_0},
\end{equation}
where we have changed to the incident angle variable $\delta$ for later comparison. 
It is interesting to note that both expressions for the angular deflection are a sum of two competing effects: $t^{(1)}_a + j^{(1)}$ in Eq.~(\ref{eq:FF}) for FF and  $\Delta^\mathrm{GH}_\mathrm{v} + \Delta^\mathrm{GH}_a$ from Eqs.~(\ref{eq:vgh}) and (\ref{eq:agh}).
In fact, on considering all prefactors the term containing $t^{(1)}_a$ is common to both formulas. 
Nevertheless the results can be quite different as shown in Fig.~\ref{fig:GenericAGH}, where the formulas for FF and the angular GH shift predict deflections in opposite directions.

%
%

\section{Discussion}
\label{sec:discussion}

In this paper we have highlighted the similarities and differences between Fresnel filtering (FF) and the angular Goos-H\"anchen shift (GH) on transmission: two related effects which give rise to a deviation from Snell's law if the incident field is not a plane wave, but a spatially localised beam of light.
Both effects are based on a modification or `filtering' of the angular spectrum, and are therefore of common in nature. 
However, different assumptions in the definition of the two effects gives rise to a difference in the magnitude of the deflections. 
To examine these differences we have extended the existing theories for FF and the angular GH shift.
For the former we have derived an analytical expression for the deviation at generic incidence, that is not for the special case of critical incidence, and for the latter we have found an approximate analytic result for the special case of critical incidence.

Before we discuss the results in detail we clearly state the differences in definition. 
FF has been defined as the difference between the peak transmission angle in far field and the angle expected from Snell's law, whereas the angular GH shift on transmission compares Snell's law with the mean angle. As a consequence FF uses the amplitude transmission coefficient, whereas the angular GH shift depends on the transmittance, that is transmission coefficient for the energy or intensity. 

In light of these differences it may seem surprising to find any similarities, but of course amplitude and intensity, near field and far field are related and we do find terms common to both formulas. 
Most prominently perhaps the logarithmic derivative of the transmission coefficient in Eqs.~(\ref{eq:agh}) and (\ref{eq:FF}), which is a familiar term from reflection.
Interestingly, it is not this shared term which explain the very similar result for critical incidence.
At this special angle the deviation from Snell's law calculated in both theories is almost identical, and the remaining difference can be explained as the difference between mean and peak angle. 
In terms of the competing terms in (\ref{eq:totalgh}) and (\ref{eq:FF}), however, it is not the common term which is dominant at the critical angle, but, somewhat surprisingly, the `virtual' shift $\Delta_\mathrm{v}^\mathrm{GH}$ or the $j^{(1)}$ term. 

For generic incidence the magnitudes of thrangular GH shift and FF can be quite different, though of course the precise difference depends on the angle of incidence. Our choice of $\theta_0$ showcases the differences.
Angles of incidence closer to critical incidence lead to smaller differences.
As the logarithmic derivative of the transmission coefficient is common to both, it is again the `virtual' term which is responsible for the difference. 
In case of the FF this term is always negative (for $n < 1$) and larger in magnitude which explains the overall negative shift.
For the angular GH shift the virtual term is also always negative if $n < 1$, but smaller than the derivative term, which is why the shift turns negative for larger angles of incidence.
The fundamental difference may be founded in the distinction between amplitude and field, which raises the question whether the difference between angular GH shift and FF has it roots in the difference between the electric field and the displacement field. 
An answer to this question, however, is outside the scope of this paper and deserves its own exposition.

\ack
We would like to thank Mark Dennis  and Alexander Ebersp\"acher for helpful discussion. JBG acknowledges support from the Royal Society via the Newton Alumnus scheme.

\end{document}